\documentclass[conference]{IEEEtran}
\IEEEoverridecommandlockouts
\usepackage{graphicx}
\usepackage{amsmath}
\usepackage{amssymb}
\usepackage{amsfonts}
\usepackage{gensymb}
\usepackage{subcaption}
\usepackage{color}
\usepackage{textcomp}
\usepackage{xcolor}
\usepackage{cite}
\usepackage{soul}

\begin{document}

\title{{Estimating the Number and Locations of Boundaries in Reverberant Environments with Deep Learning}
\thanks{{\scriptsize This work was supported by NSF grants CCF-1911094 and IIS-1730574; ONR grants N00014-18-1-2571, N00014-20-1-2534, N00014-23-1-2714, N00014-24-1-2225, N000014-23-1-2803, and MURI N00014-20-1-2787; AFOSR grant FA9550-22-1-0060; DOE grant DE-SC0020345; DOI grant 140D0423C0076; and a Vannevar Bush Faculty Fellowship, ONR grant N00014-18-1-2047.} }
}

\author{\IEEEauthorblockN{1\textsuperscript{st} Toros Arikan}
\IEEEauthorblockA{\textit{Dept. of ECE} \\
\textit{Rice University}\\
Houston, TX, USA \\
0000-0001-8748-5060}
\and
\IEEEauthorblockN{2\textsuperscript{nd} Luca M. Chackalackal}
\IEEEauthorblockA{\textit{Dept. of Computer Science} \\
\textit{Rice University}\\
Houston, TX, USA \\
0009-0008-1972-4590}
\and
\IEEEauthorblockN{3\textsuperscript{rd} Fatima Ahsan}
\IEEEauthorblockA{\textit{Dept. of ECE} \\
\textit{Rice University}\\
Houston, TX, USA \\
0000-0001-5425-9040}
\and
\IEEEauthorblockN{4\textsuperscript{th} Konrad Tittel}
\IEEEauthorblockA{\textit{Dept. of ECE} \\
\textit{Rice University}\\
Houston, TX, USA \\
0009-0009-5105-8716} 
\and
\IEEEauthorblockN{5\textsuperscript{th} Andrew C. Singer}
\IEEEauthorblockA{\textit{Dept. of ECE} \\
\textit{Stony Brook University}\\
Stony Brook, NY, USA \\
0000-0001-9926-7036}
\and
\IEEEauthorblockN{6\textsuperscript{th} Gregory W. Wornell}
\IEEEauthorblockA{\textit{Dept. of EECS} \\
\textit{Massachusetts Institute of Technology}\\
Cambridge, MA, USA \\
0000-0001-9166-4758}
\and
\IEEEauthorblockN{7\textsuperscript{th} Richard G. Baraniuk}
\IEEEauthorblockA{\textit{Dept. of ECE} \\
\textit{Rice University}\\
Houston, TX, USA \\
0000-0002-0721-8999}
}



\maketitle

\begin{abstract}
Underwater acoustic environment estimation is a challenging but important task for remote sensing scenarios. Current estimation methods require high signal strength and a solution to the fragile echo labeling problem to be effective. In previous publications, we proposed a general deep learning-based method for two-dimensional environment estimation which outperformed the state-of-the-art, both in simulation and in real-life experimental settings. A limitation of this method was that some prior information had to be provided by the user on the number and locations of the reflective boundaries, and that its neural networks had to be re-trained accordingly for different environments. Utilizing more advanced neural network and time delay estimation techniques, the proposed improved method no longer requires prior knowledge the number of boundaries or their locations, and is able to estimate two-dimensional environments with one or two boundaries. Future work will extend the proposed method to more boundaries and larger-scale environments.
\end{abstract}

\section{Introduction}
Estimation of the reflective boundaries in reverberant environments is an important yet difficult task for underwater and indoor acoustics \cite{gerstoft1,gerstoft2,gerstoft_indoor}, that allows the use of non-line of sight (NLOS) arrivals to enhance localization performance \cite{simultaneous_tx_boundary_slam}. Over short ranges, these boundaries can be approximated as planar, which yield mirror images of the true emitter as `virtual' emitters. Thus, given a known or estimated emitter location, and known receiver locations, a variety of methods can be used for boundary estimation through virtual emitter localization \cite{LS_alg}, \cite{EDM_boundaries}, \cite{convex_opt_bounds_known}. However, low signal-to-noise ratios (SNRs) \cite{high_snr_impractical} are not addressed by these existing methods. They also generally require the solution of a difficult combinatorial echo labeling problem to differentiate between the boundaries \cite{echo_labeling}, complicated by missing or spurious echoes.

\begin{figure}[t]
\centering
\includegraphics[width=0.25\textwidth]{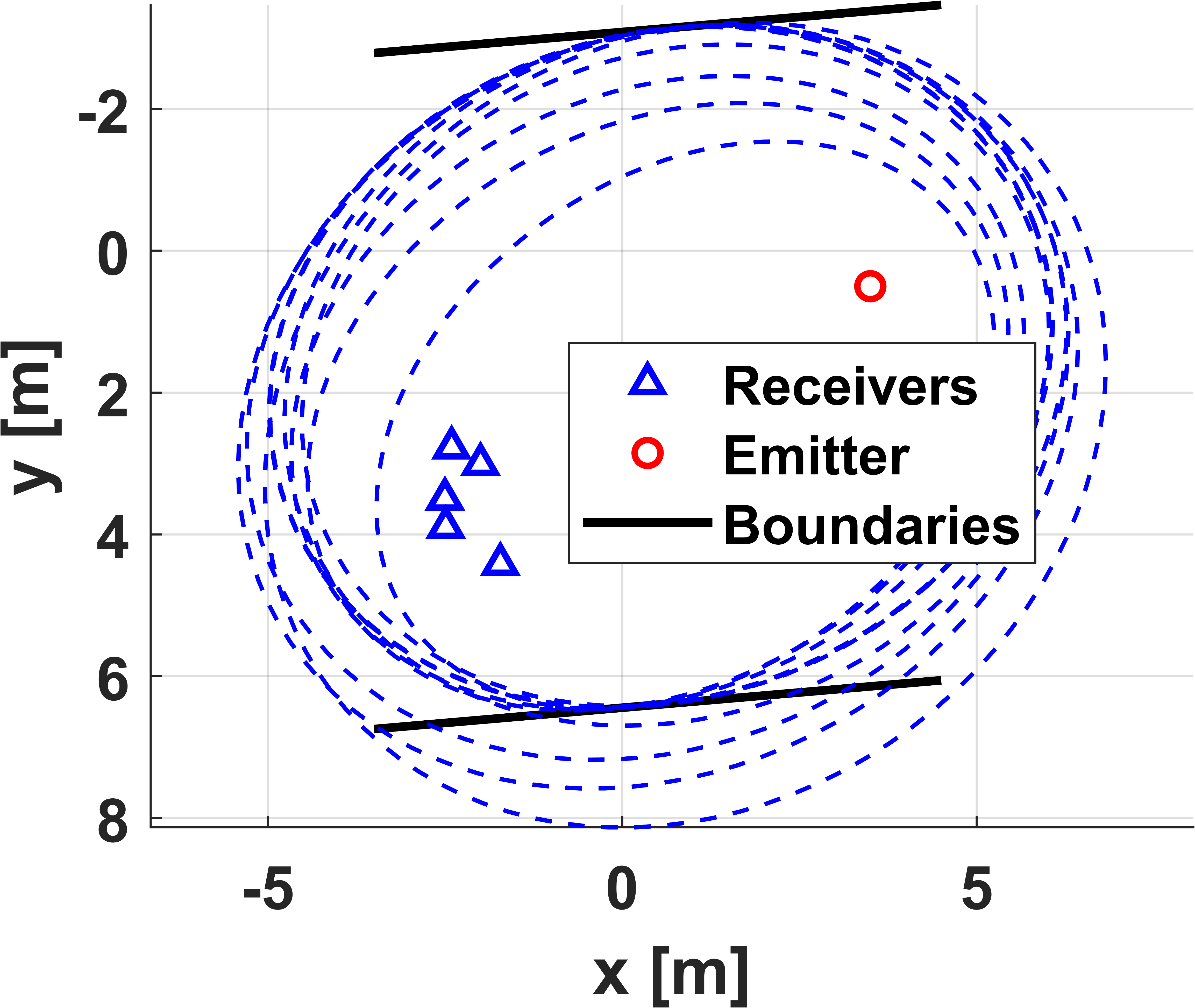}\vspace{-0.25cm}
\caption{General simulation setting: each NLOS arrival yields an ellipse whose common tangents are reflective boundaries.}
\label{fig:underwater_environment}
\end{figure}

In previous work \cite{cotans_nn}, \cite{neuro_cotans}, we introduced the convolutional neural network-based (CNN) Neuro-COTANS method for boundary estimation, to overcome these challenges. Neuro-COTANS leveraged the fact that in 2D, a NLOS arrival corresponds to an ellipse with a path distance of $d_{\textrm{NLOS}}$, whose foci are the emitter and receiver locations. Multiple receivers define multiple ellipses whose common tangents are the boundaries, as in Fig.~\ref{fig:underwater_environment}. We parametrized the tangents to ellipses by their range $\rho$ and azimuth $\theta$ \cite{simultaneous_tx_boundary_slam}, calling this $\rho$- and $\theta$-space the common tangents to spheroids (COTANS) domain. Creating COTANS-domain images where each ellipse corresponded to a curve as in Fig.~\ref{fig:two_boundary_cotans}, we trained an AlexNet \cite{alexnet} neural network (NN) to estimate boundaries from the unlabeled NLOS distances. Neuro-COTANS proved more robust than methods which applied heuristic smoothing filters to COTANS images \cite{hough_first_paper}, \cite{hough_second_paper}, \cite{hough_echo_labeling}. It also outperformed the state-of-the-art least-squares (LS) \cite{LS_alg} and Euclidean distance matrices (EDM) \cite{EDM_review}, \cite{EDM_boundaries} methods by up to 6 dB SNR, even when they were advantageously initialized with the correct echoes.

While Neuro-COTANS achieved groundbreaking performance, it had several  shortcomings that motivated future work. One of them was that the number of boundaries had to be assumed known. Another was that the boundaries were assumed to be known to within $10^{\circ}$ in azimuth, informed by prior knowledge. These constraints meant that the NN had to be retrained for different environments. 

In this paper, we introduce U-COTANS, a U-Net \cite{unet_original} method which introduces the critical operational capabilities of estimating the number of boundaries and covering the entire COTANS image of a given resolution and scale. We retain the multi-scale, multi-stage framework of Neuro-COTANS \cite{neuro_cotans}, while exploiting the proven capabilities of U-Nets to solve difficult 2D estimation problems through image segmentation \cite{fatima_unet_thesis}. Through a careful choice of training and test models, we maintain our groundbreaking performance while incorporating new abilities into our overall approach.

The paper is organized as follows. In Section~\ref{sec:signal_model}, we briefly review the COTANS image generation framework introduced in our past work. The U-COTANS method is detailed in Section~\ref{sec:cotans_overview}. Simulation results are presented in Section~\ref{sec:results}, and final remarks are given in Section~\ref{sec:conclusion}. Our code is publicly available at \textcolor{blue}{\underline{https://github.com/torosarikan/U-COTANS}}.
\section{Problem Formulation} \label{sec:signal_model}
\begin{figure}[tbp]
\centering
\includegraphics[width=0.35\textwidth]{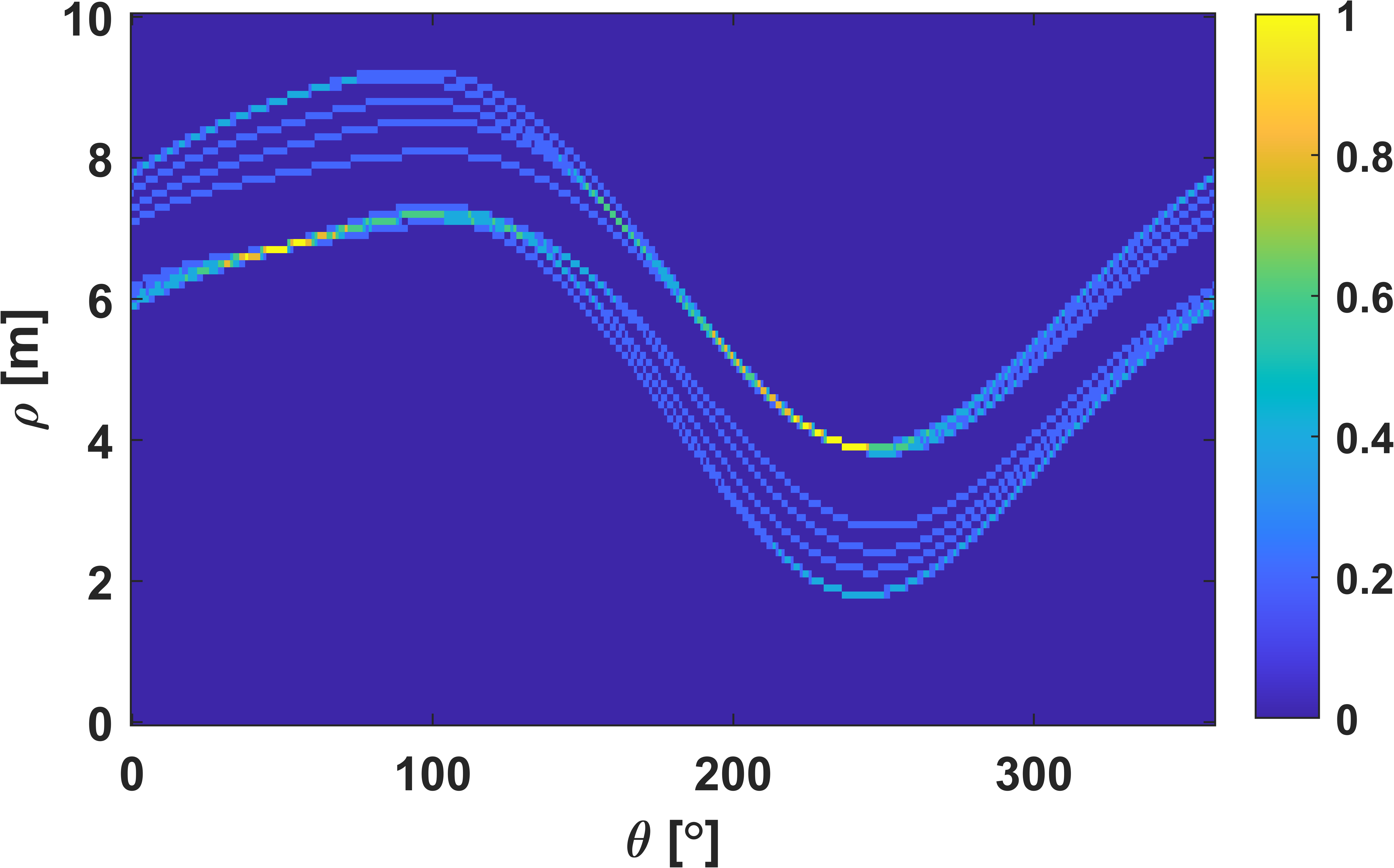}
\caption{A high-SNR COTANS image with two boundaries, with the curves from the respective boundaries properly intersecting at the ground-truth $(\rho,\theta)$ values.}
\label{fig:two_boundary_cotans}
\end{figure}

We continue to use the static 2D environment model that was presented in prior work \cite{cotans_nn}, \cite{neuro_cotans}. There are $N$ planar boundaries in the environment, as described by the range $\rho\in\mathbb{R}_+$ and azimuth $\theta\in[-\pi,\pi)$ of their normal vector relative to the (arbitrarily-chosen) origin. There is a single isotropic emitter in the environment at a known location $\mathbf{p}_{\textrm{e}}=\left[x_{\textrm{e}}\;y_{\textrm{e}}\right]^{\rm{T}}$, and $M$ isotropic receivers at known locations $\mathbf{p}_{\textrm{r},i}=\left[x_{i}\;y_{i}\right]^{\rm{T}}$. The speed of sound is assumed to be constant \cite{snapping_shrimp}. Given the energy of a received pulse as $E_r$, the SNR is defined as $E_r/N_0$, where $N_0$ is the one-sided power spectral density of the Gaussian noise. At high SNR, the error in the time delay estimates $\{\hat{\tau}_{i,j}\}$ are Gaussian \cite{range_estimation_textbook}, but there is an SNR threshold below which large `global errors' occur \cite{time_delay_bounds}. The COTANS transform \cite{neuro_cotans} conceptualizes a boundary that is defined by $\rho$ and $\theta$ as a point $(\rho,\theta)$ in a COTANS domain, and calculates the function $\rho(\theta)$ for each of $\{\hat{\tau}_{i,j}\}$ \cite{3d_hough_transform}. Summing the resulting curves in a discretized $\rho\times\theta$ space yields a COTANS image as in Fig.~\ref{fig:two_boundary_cotans}, with maxima at the true boundaries $\{(\rho_{j},\theta_{j})\}$ at high SNR.

In Neuro-COTANS \cite{neuro_cotans}, we used matched filtering (MF) to produce the $\{\hat{\tau}_{i,j}\}$. While this was adequate for prototyping, there are some fundamental issues with using MF for multipath time delay estimation. In multipath environments, there is the likelihood of overlap of different arrivals. While MF is the maximum likelihood (ML) estimator for line of sight (LOS) arrivals in Gaussian noise, its performance deteriorates if the MF results of different multipath arrivals overlap. In light of this difficulty, we instead adopt the space-alternating generalized expectation-maximization (SAGE) algorithm for the estimation of time delays of potentially overlapping arrivals \cite{sage_toa_est}. SAGE is an extension of expectation-maximization (EM) techniques \cite{em_toa}, and is designed to prevent convergence to a single copy of the emitted signal. Although omitted for brevity, SAGE does indeed yield improved estimates $\{\hat{\tau}_{i,j}\}$ on our datasets.

\section{U-COTANS for Boundary Estimation} \label{sec:cotans_overview}
In Neuro-COTANS \cite{neuro_cotans}, the outputs of the NN were the boundary parameter estimates $[\hat{\rho}_{1}\cdots\hat{\rho}_{N}\:\hat{\theta}_{1}\cdots\hat{\theta}_{N}]^{\textrm{T}}$, each scaled to a range of $[0,\,1]$ by dividing each $\rho$ by a $\rho_{\textrm{max}}$ (10 m in our case), and each $\theta$ by 360\textdegree. We used the correct $\left[\rho_{1}\cdots\rho_{N}\:\theta_{1}\cdots\theta_{N}\right]^{\textrm{T}}$ for training outputs, and training input images as in Fig.~\ref{fig:two_boundary_cotans} were generated by simulating scenarios with randomized $\mathbf{p}_{\textrm{e}}$ and $\{\mathbf{p}_{\textrm{r},i}\}$, as in Fig.~\ref{fig:underwater_environment}. 

\begin{figure}[tbp]
\centering
\includegraphics[width=0.35\textwidth]{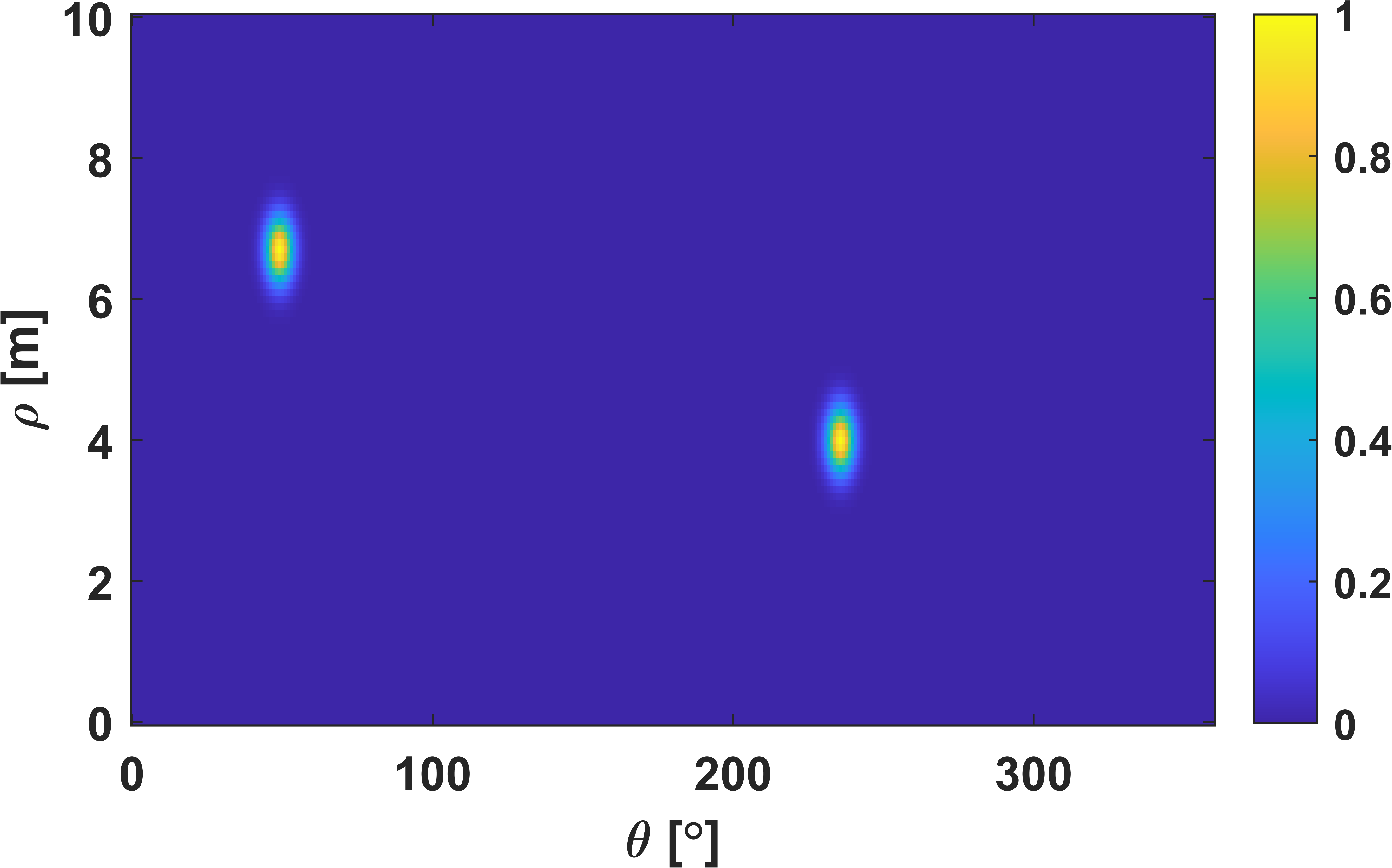}
\caption{The BEI corresponding to the COTANS image in Fig.~\ref{fig:two_boundary_cotans}, with Gaussian pulses superimposed over the ground-truth $(\rho,\theta)$ values.}
\label{fig:two_boundary_bei}
\end{figure}

Whereas Neuro-COTANS re-purposed the 8-layer AlexNet architecture by replacing the classification layer with a regression layer \cite{alexnet_regression}, U-COTANS retains the established structure of a U-Net \cite{unet_original}, while modifying some of its hyperparameters. Specifically, we set the number of channels to 1, the learning rate to 0.05, and the weight decay to 0. The fundamental innovation in U-COTANS is the use of an image segmentation methodology, rather than a regression one. The inputs are COTANS images as in Fig.~\ref{fig:two_boundary_cotans}, and the training outputs are images with the same dimensions (e.g., $101\times360$), but with pulses overlaid on the correct COTANS-domain boundary locations. Specifically, each ground truth boundary location is made the center of a 2D Gaussian pulse, truncated to a square 10 pixels wide, as in Fig.~\ref{fig:two_boundary_bei}. We call this a boundary estimate image (BEI). 

The underlying idea of U-COTANS is that the BEI approximates a heatmap of the likelihood of true COTANS-domain locations. Excepting global errors, COTANS curves either intersect at a point and yield a pixel of high intensity, or are superposed in the vicinity of a point and produce a region of high intensity. The guiding observation of Neuro-COTANS was that a regression NN could learn a high-level estimator which could leverage this local intensity and other global information for accurate inference. U-COTANS makes this idea explicit: a region of high intensity pixels will correspond to a high-intensity region in the NN output, with the rest of the image effectively having been thresholded.

To estimate the boundaries using a BEI that is produced from a COTANS image by U-COTANS, we find the maximum of the BEI (indicating the presence of a pulse); then to calculate the center of mass of the pixels around this maximum (returning a potentially improved boundary estimate); and finally to zero out the vicinity of this maximum and find a new maximum in the BEI (moving to another pulse). 

\subsection{Improved environment generalization}

The segmentation-based U-Net approach of U-COTANS offers critical capability advantages over the regression-based AlexNet approach of Neuro-COTANS. When a NN is trained to estimate boundaries that only come from a specific $(\rho,\theta)$ region, the resulting NN estimator limits its predictions to this constrained region. This is advantageous when prior knowledge is available, leveraging this information to improve performance. As the permissible range of $(\rho,\theta)$ grows to encompass the entire COTANS image for a given scale, however, the probability of any single pixel being selected is reduced, and even a large training dataset will fail to generalize to all possible environments. 

In contrast, a BEI such as Fig.~\ref{fig:two_boundary_bei} is centered on the boundary estimate, but also has positive values for the surrounding pixels within a radius around this estimate. Thus, a training dataset is able to provide a covering of the entire decision space, overcoming the generalization hurdle.

\subsection{Estimation of the number of boundaries}

\begin{figure}[tbp]
\centering
\includegraphics[width=0.35\textwidth]{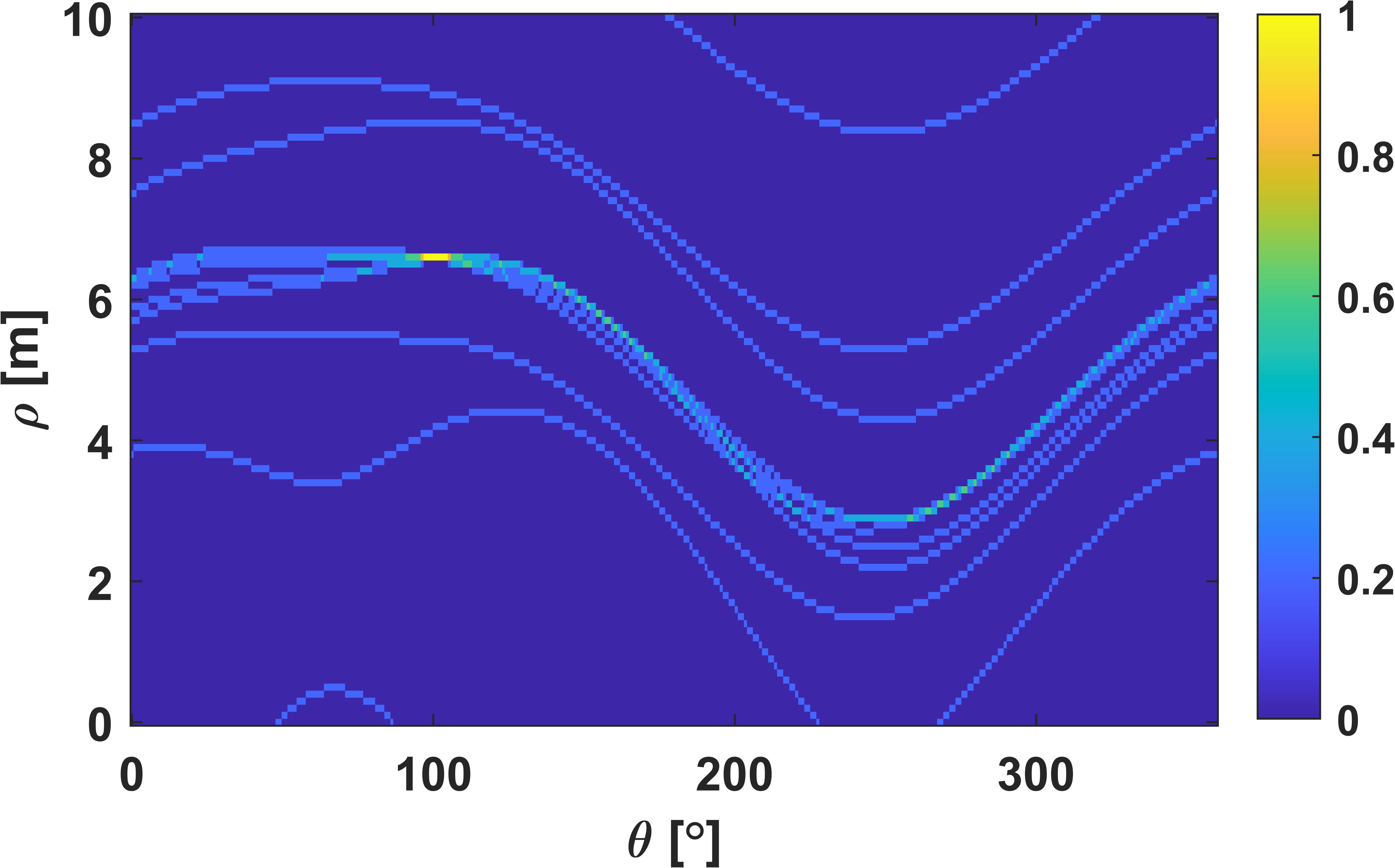}
\caption{A COTANS image with $N=1$, with one set of curves properly intersecting at the ground-truth $(\rho,\theta)$ value, and the other set observed as global errors.}
\label{fig:one_boundary_cotans}
\end{figure}

\begin{figure}[tbp]
\centering
\includegraphics[width=0.35\textwidth]{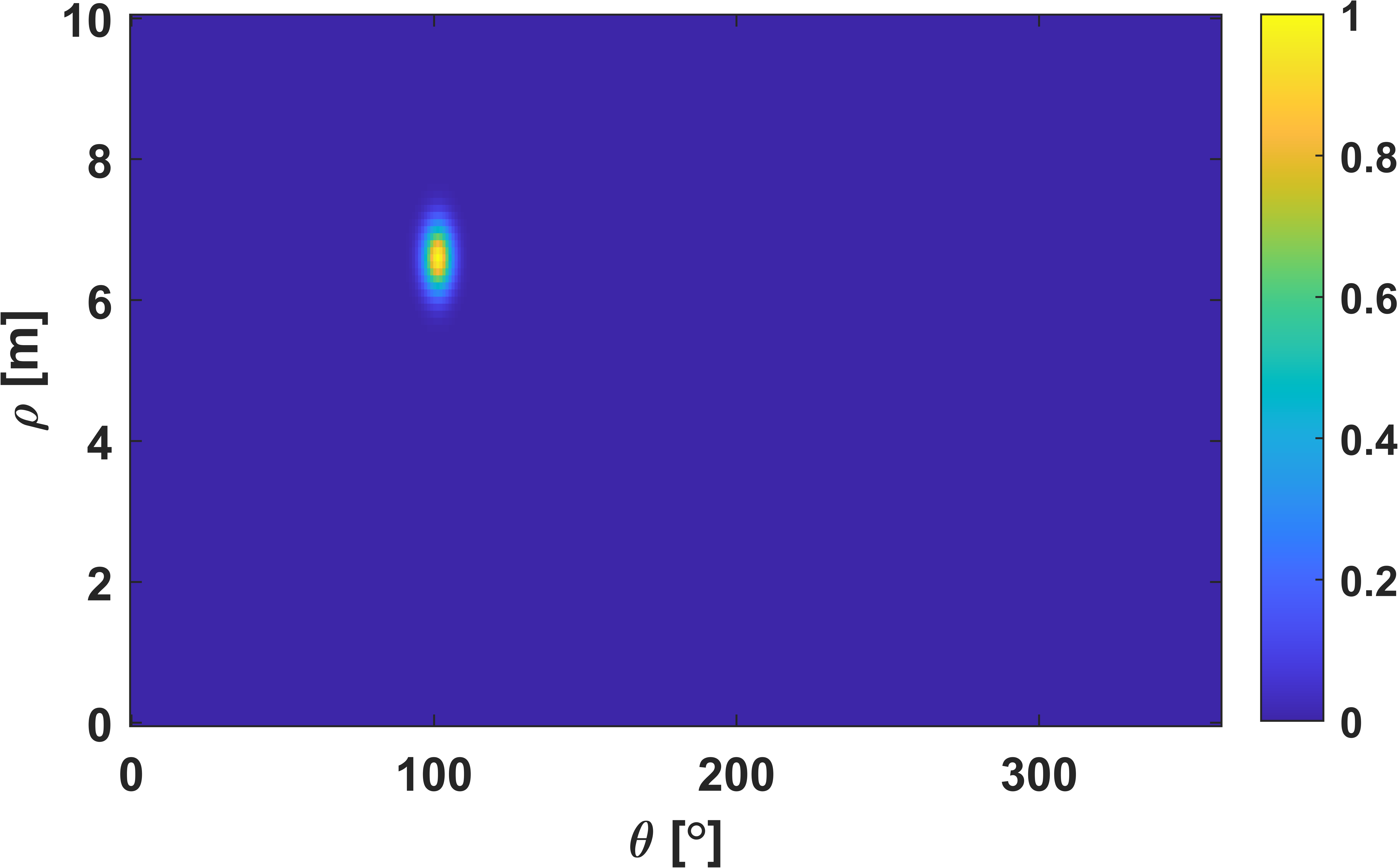}
\caption{The BEI of the COTANS image in Fig.~\ref{fig:one_boundary_cotans}, with a Gaussian pulse superimposed over the ground-truth $(\rho,\theta)$.}
\label{fig:one_boundary_bei}
\end{figure}

U-COTANS also estimates the number of boundaries $N$ in the environment. In Neuro-COTANS, $N$ was assumed known since this dictated the size of the last regression layer of the network. U-COTANS instead allows us to use different BEIs to estimate $N$. Consider the case of only one boundary being present in the environment, and that we know that $N$ is a maximum of two. Then, when we obtain time delay estimates assuming $N=2$ as in Fig.~\ref{fig:one_boundary_cotans}, only one set of COTANS curves will result from the true multipath signal. The other set of curves will instead be produced by random noise peaks, and are unlikely to intersect at all. Therefore, we also train U-COTANS with BEIs that can have only one pulse, as in Figs.~\ref{fig:one_boundary_bei}, so that the NN outputs not just the positions of the boundaries but their number as well.

\section{Simulation Results}\label{sec:results}

We first present the performance of U-COTANS in a simulation setting that is a more generalized case of a two-boundary shallow-water underwater acoustic channel, and compare it to the LS algorithm. We then present our boundary number estimation results for the same environments with one or two boundaries present.

We train U-COTANS on 9 SNR levels, equally spaced in the 13 to 21 dB SNR range  (covering the medium- and high-SNR regimes), generating 50,000 training, 10,000 validation, and 50,000 test images per SNR. The $\theta$-parameter of each boundary is picked with equal probability to lie in one of the four quadrants, with the additional constraint that the two boundaries are separated by a minimum of $30^{\circ}$ (to match real-life environments). Within each quadrant, $\theta$ for that boundary is a uniformly distributed random variable. The $\rho$-parameter is uniformly distributed up to 8 m in each quadrant, with a minimum $\rho$ enforced with each quadrant to prevent a boundary coming in between the receivers and the emitter (e.g., $\rho > 3$ m in quadrant 3 for our particular simulation scenario). The $\mathbf{p}_{\textrm{e}}$ and $\{\mathbf{p}_{\textrm{r},i}\}$ are drawn from a uniform distribution over 2 meter-wide areas centered on the points $(3.5,\,0.5)$ m and $(-2.5,\,3.5)$ m, respectively. Our performance metric is the range RMSE (in m) over all $N$ boundaries and $K$ environment realizations for each SNR $S$:
\begin{equation}
    \rho_{\textrm{RMSE}}(S) \triangleq \sqrt{\frac{\sum_{j=1}^{N}\sum_{k=1}^{K}\left(\rho_{j,k}^{(S)}-\hat{\rho}_{j,k}^{(S)}\right)^{2}}{NK}}.
\end{equation}
The resulting performance curves for $\theta$ are qualitatively similar to the performances for $\rho$ presented here. 

\begin{figure}[tbp]
\centering
\includegraphics[width=6cm]{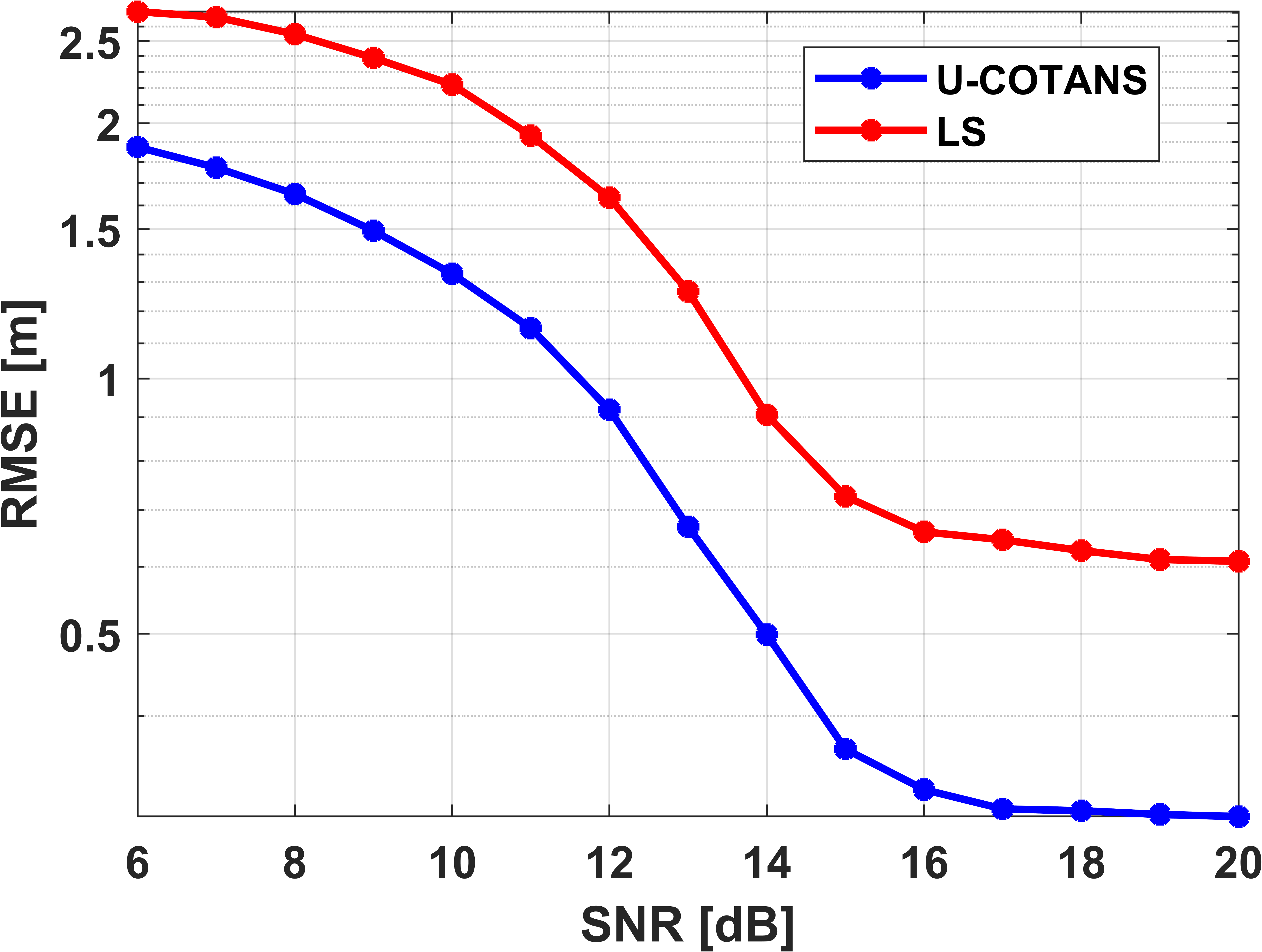}
\caption{Range RMSE vs.\ SNR of U-COTANS and LS.}
\label{fig:two_bound_rmse}
\end{figure}

We compare U-COTANS to LS that has been given the advantage of initialization at the ground truth virtual emitter locations and with the correct echo labeling. Fig.~\ref{fig:two_bound_rmse} shows the average range RMSE of these methods vs.\ the SNR. U-COTANS outperforms LS by a minimum of 3 dB SNR, and performs much better in the low and high SNR regimes. The leveling off in performance in both methods is due to specific challenging time delay estimation scenarios where a more advanced time delay estimation front-end may be needed for both algorithms. Recall that since LS requires echo labeling and random initialization that in practice can also cause large errors, its real-life performance will be worse than in Fig.~\ref{fig:two_bound_rmse}.

\begin{figure}[tbp]
\centering
\includegraphics[width=5cm]{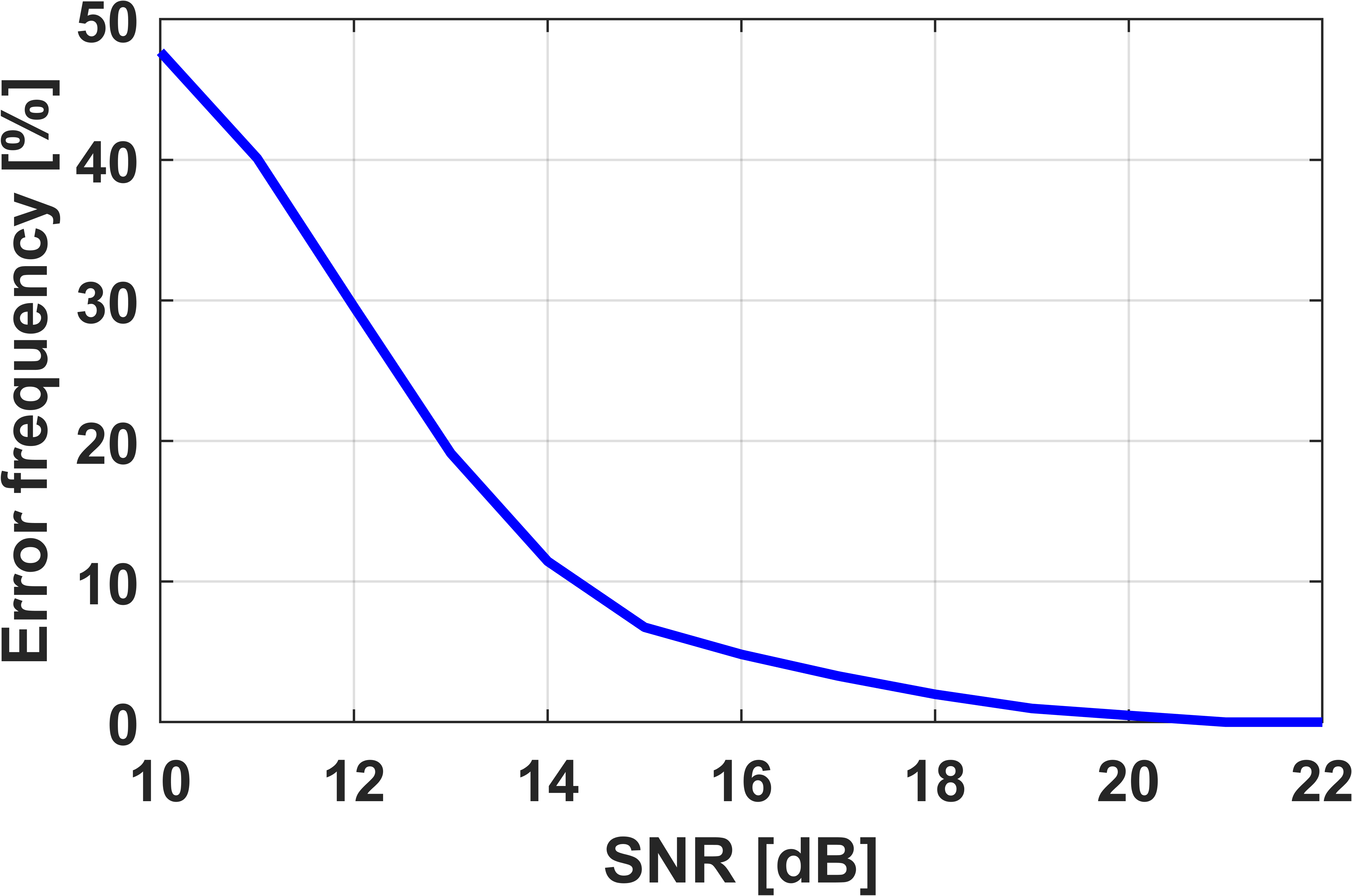}
\caption{Accuracy of estimating the number of boundaries $N$ in the environment with U-COTANS.}
\label{fig:numbounds_estimation}
\end{figure}

For the U-COTANS boundary estimation trials, we generate the simulated environments with a 50\% chance of having $N=1$. Our performance metric is the frequency of inaccurate $N$ estimates, as a percentage of the number of trials. Our results in Fig.~\ref{fig:numbounds_estimation} demonstrate strong performance at medium to high SNR for this novel capability, with perfect results being achieved at high SNR.  We observe that the performance in Fig.~\ref{fig:numbounds_estimation} closely mirrors the performance curve in Fig.~\ref{fig:two_bound_rmse}, suggesting that the detection task of choosing $N$ and the estimation task of determining $(\rho,\theta)$ for a given $N$ are fundamentally related. In a future study, we will investigate the correspondence between the trials with high range RMSE and the trials where $N$ is estimated incorrectly.

In a separate set of experiments omitted here for brevity, we trained U-COTANS on environments where the boundaries were known to within $10^{\circ}$ in azimuth, and compared its performance to our past Neuro-COTANS performance \cite{neuro_cotans}. We observed that U-COTANS reproduced the performance of Neuro-COTANS, performing slightly better when using SAGE instead of MF. Although it is intuitive that two CNN methods trained on the same dataset with the same error metric would yield similar performances, the segmentation and regression approaches are fundamentally different methodologies for approaching the same task. Hence, the two NNs seem to be learning the same correct underlying global optimization algorithm for this task.

\section{Concluding Remarks} \label{sec:conclusion}
We propose the U-COTANS method for 2D boundary estimation, as a major enhancement of our previously published Neuro-COTANS methodology. U-COTANS continues to deliver robust performance that is superior to the state-of-the-art alternatives such as LS. Additionally, U-COTANS no longer needs to be re-trained for different environments and can handle \emph{any} environment of a given COTANS image scale. U-COTANS also introduces the capability of directly estimating the number of boundaries in a given environment from the NN results instead of requiring other sensors or estimation front-ends, which to the best of our knowledge is not shared by \emph{any} state-of-the-art boundary estimation methods. The results that we have achieved bring us close to a plug-and-play environment estimation method that can handle general 2D environments over any SNR range and any reasonable number of boundaries.

Future work will focus in the near term on applying U-COTANS to more than 2 boundaries and over longer distances of up to 1 km; fundamentally, this is a straightforward extension of the method to several additional zoomed stages, and of re-training on additional environment geometries. Another operationally important extension will be the case of a moving emitter, which will allow for higher performance over time due to the spatial diversity provided by the different emitter locations. This will lead to higher performance over time as compared to single-snapshot environment estimation. Finally, while U-COTANS works with a 2D setting, its operation will ultimately be extended to 3D. 

U-COTANS has some key operational advantages for estimating $N$ (i.e., the number of copies of the emitted signal in the received signals) over the established information-theoretic methods such as the Akaike information criterion (AIC) \cite{akaike_original} and the minimum description length (MDL) \cite{bresler_mdl}, for this particular environment estimation problem. These alternative methods are biased towards producing a larger estimate for $N$ when the signal observation window is longer, while our method has no such limitations. Our method is also able to leverage, not just the received signals as is the case for AIC and MDL, but the geometric setting information encoded in the COTANS image and the accuracy of the resulting boundary estimates as a metric. In future work, we will rigorously compare the performances of AIC and MDL to that of U-COTANS.


\begin{thebibliography}{10}

\bibitem{gerstoft1}
H. Niu, E. Ozanich, and P. Gerstoft,
\newblock ``Ship localization in {S}anta {B}arbara {C}hannel using machine learning classifiers,''
\newblock {\em J. Acoust. Soc. Am.}, vol. 142, no. 5, pp. 455--460, 2017.

\bibitem{gerstoft2}
H. Niu, Z. Gong, E. Ozanich, P. Gerstoft, H. Wang, and Z. Li,
\newblock ``Deep-learning source localization using multi-frequency magnitude-only data,''
\newblock {\em J. Acoust. Soc. Am.}, vol. 146, no. 1, pp. 211--222, 2019.

\bibitem{gerstoft_indoor}
Y. Wu, R. Ayyalasomayajula, M.~J. Bianco, D. Bharadia, and P. Gerstoft,
\newblock ``Sslide: {S}ound source localization for indoors based on deep learning,''
\newblock in {\em {ICASSP} 2021-2021 {IEEE} {I}nt. Conf. on Acoustics, Speech and Signal Processing}, Jun 2021, pp. 4680--4684.

\bibitem{simultaneous_tx_boundary_slam}
H. Naseri and V. Koivunen,
\newblock ``Cooperative simultaneous localization and mapping by exploiting multipath propagation,''
\newblock {\em {IEEE} Signal Process. Mag.}, vol. 65, no. 1, pp. 200--211, 2016.

\bibitem{LS_alg}
K.~W. Cheung, H.~C. So, W.~K. Ma, and Y.~T. Chan,
\newblock ``{Least squares algorithms for time-of-arrival-based mobile location},''
\newblock {\em {IEEE} Trans. Signal Process.}, vol. 52, no. 4, pp. 1121--113, 2004.

\bibitem{EDM_boundaries}
I. Dokmanic, R. Parhizkar, A. Walther, Y.~M. Lu, and M. Vetterli,
\newblock ``{Acoustic echoes reveal room shape},''
\newblock {\em Proc. Nat. Acad. Sciences}, vol. 110, no. 30, pp. 12186--12191, 2013.

\bibitem{convex_opt_bounds_known}
H. Naseri, M. Costa, and V. Koivunen,
\newblock ``Multipath-aided cooperative network localization using convex optimization,''
\newblock in {\em 48th Asilomar Conf. Signals, Syst. Comput.}, 2014, pp. 1515--1520.

\bibitem{high_snr_impractical}
D. Dardari, A. Conti, U. Ferner, A. Giorgetti, and M.~Z. Win,
\newblock ``Ranging with ultrawide bandwidth signals in multipath environments,''
\newblock {\em Proc. {IEEE}}, vol. 97, no. 2, pp. 404--426, 2000.

\bibitem{echo_labeling}
M. Crocco, A. Trucco, and A.~D. Bue,
\newblock ``Uncalibrated 3{D} room geometry estimation from sound impulse responses,''
\newblock {\em J. Franklin Institute}, vol. 354, no. 18, pp. 8678--8709, 2017.

\bibitem{cotans_nn}
T. Arikan, A. Weiss, H. Vishnu, G.~B. Deane, A.~C. Singer, and G.~W. Wornell,
\newblock ``Learning environmental structure using acoustic probes with a deep neural network,''
\newblock in {\em {ICASSP} 2023-IEEE Int. Conf. Acoustics, Speech and Signal Process.}, 2023, vol.~26, pp. 1--5.

\bibitem{neuro_cotans}
T. Arikan, A. Weiss, H. Vishnu, G.~B. Deane, A.~C. Singer, and G.~W. Wornell,
\newblock ``A deep learning method for reflective boundary estimation,''
\newblock {\em J. Acoust. Soc. Am.}, vol. 156, no. 1, pp. 65--80, 2024.

\bibitem{alexnet}
A. Krizhevsky, I. Sutskever, and G.~E. Hinton,
\newblock ``Image{N}et classification with deep convolutional neural networks,''
\newblock in {\em Advances in neural inf. process. syst.}, 2012, vol.~25.

\bibitem{hough_first_paper}
F. Antonacci, A. Sarti, and S. Tubaro,
\newblock ``Geometric reconstruction of the environment from its response to multiple acoustic emissions,''
\newblock in {\em IEEE Int. Conf. on Acoust., Speech, Signal Process.}, 2010, pp. 2822--2825.

\bibitem{hough_second_paper}
F. Antonacci, J. Filos, M.~R. Thomas, E.~A. Habets, A. Sarti, P.~A. Naylor, and S. Tubaro,
\newblock ``Inference of room geometry from acoustic impulse responses,''
\newblock {\em IEEE Trans. Audio, Speech, Lang. Process.}, vol. 20, no. 10, pp. 2683--2695, 2012.

\bibitem{hough_echo_labeling}
S. Park and J. Choi,
\newblock ``Iterative echo labeling algorithm with convex hull expansion for room geometry estimation,''
\newblock {\em IEEE/ACM Trans. Audio, Speech, Lang. Process.}, vol. 29, no. 3, pp. 1463--1478, 2021.

\bibitem{EDM_review}
I. Dokmanic, R. Parhizkar, J. Ranieri, and M. Vetterli,
\newblock ``{Euclidean Distance Matrices: Essential theory, algorithms, and applications},''
\newblock {\em {IEEE} Signal Process. Mag.}, vol. 32, no. 6, pp. 12--30, Nov 2015.

\bibitem{unet_original}
O. Ronneberger, P. Fischer, and T. Brox,
\newblock ``U-net: {C}onvolutional networks for biomedical image segmentation,''
\newblock in {\em Medical image computing and computer-assisted intervention–{MICCAI} 2015}. 2015, vol.~26, pp. 234--241, pringer International Publishing.

\bibitem{fatima_unet_thesis}
F. Ahsan,
\newblock {\em {EMvelop Stimulation: Minimally Invasive Deep Brain Stimulation using Temporally Interfering Electromagnetic Waves}},
\newblock Ph.D. thesis, Rice University, 2024.

\bibitem{snapping_shrimp}
Y.~M. Too, M. Chitre, G. Barbastathis, and V. Pallayil,
\newblock ``Localizing snapping shrimp noise using a small-aperture array,''
\newblock {\em {IEEE} J. Oceanic Eng.}, vol. 44, no. 1, pp. 207--219, 2017.

\bibitem{range_estimation_textbook}
C. Cook,
\newblock {\em Radar signals: {A}n introduction to theory and application},
\newblock Elsevier, 2012.

\bibitem{time_delay_bounds}
D. Dardari, C. Chong, and M. Win,
\newblock ``Improved lower bounds on time-of-arrival estimation error in realistic {UWB} channels,''
\newblock in {\em 2006 IEEE Int. Conf. Ultra-Wideband}, 2006, pp. 531--537.

\bibitem{3d_hough_transform}
D. Borrmann, J. Elseberg, K. Lingemann, and A. Nüchter,
\newblock ``The 3{D} {H}ough transform for plane detection in point clouds: {A} review and a new accumulator design,''
\newblock {\em 3{D} Research}, vol. 2, no. 2, pp. 3, 2011.

\bibitem{sage_toa_est}
R. Demirli and J. Saniie,
\newblock ``Model-based estimation of ultrasonic echoes. {P}art {I}: {A}nalysis and algorithms,''
\newblock {\em {IEEE} Trans. Ultrason., Ferroelect., Freq. Contr.}, vol. 48, no. 3, pp. 787--802, 2001.

\bibitem{em_toa}
M. Feder and E. Weinstein,
\newblock ``Parameter estimation of superimposed signals using the {EM} algorithm,''
\newblock {\em {IEEE} Trans. Acoust., Speech, Signal Process.}, vol. 36, no. 4, pp. 477--489, 1988.

\bibitem{alexnet_regression}
C. Szegedy, A. Toshev, and D. Erhan,
\newblock ``Deep neural networks for object detection,''
\newblock in {\em Advances in neural inf. process. syst.}, 2013, vol.~26.

\bibitem{akaike_original}
H. Akaike,
\newblock ``{A new look at the statistical model identification},''
\newblock {\em {IEEE} Trans. Automat. Contr.}, vol. 19, no. 6, pp. 716--723, 1974.

\bibitem{bresler_mdl}
S.~F. Yau and Y. Bresler,
\newblock ``{Maximum likelihood parameter estimation of superimposed signals by dynamic programming},''
\newblock {\em {IEEE} Trans. Signal Process.}, vol. 41, no. 2, pp. 804--820, 1993.

\end{thebibliography}

\end{document}